# Emerging Search Regimes:

# Measuring Co-evolutions among Research, Science, and Society

Gaston Heimeriks [a] and Loet Leydesdorff [b]

[a] *Innovation Studies, Copernicus Institute, Utrecht University, Utrecht, The Netherlands*

[b] *Amsterdam School of Communications Research (ASCoR), University of Amsterdam, Amsterdam, The Netherlands*

Corresponding author: Gaston Heimeriks
Email: gheimeriks@gmail.com
Innovation Studies, Copernicus Institute, Utrecht University
Heidelberglaan 2, 3584 CS Utrecht, The Netherlands

Scientometric data is used to investigate empirically the emergence of search regimes in Biotechnology, Genomics, and Nanotechnology. Complex regimes can emerge when three independent sources of variance interact. In our model, researchers can be considered as the nodes that carry the science system. Research is geographically situated with site-specific skills, tacit knowledge and infrastructures. Second, the emergent science level refers to the formal communication of codified knowledge published in journals. Third, the socio-economic dynamics indicate the ways in which knowledge production relates to society. Although Biotechnology, Genomics, and Nanotechnology can all be characterised by rapid growth and divergent dynamics, the regimes differ in terms of self-organization among these three sources of variance. The scope of opportunities for researchers to contribute within the constraints of the existing body of knowledge are different in each field. Furthermore, the relevance of the context of application contributes to the knowledge dynamics to various degrees.

**Introduction**

Knowledge is increasingly recognised as a driver of productivity and economic growth, as well as a vital resource in addressing societal challenges. This leads to increased (policy-) attention to the role of knowledge in societal and economic performance. The term 'knowledge-based economy' stems from this fuller recognition of the place of organised knowledge and technology in modern societies (OECD 1996). However, the dynamics of codified knowledge are a complex matter.

Recent studies on search regimes show that different scientific fields may exhibit very different dynamics while co-evolving with socio-economic environments. Following Bonaccorsi (2008), we consider search regimes as a set of broad dimensions that capture the knowledge dynamics in different fields. Thus, 'search regime' is a summary description of the pattern of development of scientific knowledge and of the actual carrying out of scientific research. For example, scientific fields can be shown to differ in their rate of growth, the degree of divergence, and the level of complementarity.

Copying best practices in research and innovation policy, as identified by benchmarking studies, may be popular amongst policy makers, but can be expected to fail because of the differences among fields of knowledge production (Asheim et al. 2006). Research fields accordingly exhibit institutional and localized knowledge dynamics that may respond differently to government interventions. How can one understand the different search regimes in terms of relevant co-evolutions?

We aim at providing a co-evolutionary model for knowledge dynamics in different search regimes by elaborating the notion of science as a complex adaptive system that allows for empirical operationalisation. Numerous existing studies have focused on different dynamics of knowledge production (e.g., Gibbons et al. 1994; Whitley 2000). However, these studies have several shortcomings. Firstly, only limited empirical work has been carried out to support claims of different modes of knowledge dynamics. Furthermore, previous strategies to understand science often have focused mainly on only a single evolutionary context.

The micro-level of knowledge production (context of discovery) was addressed in laboratory studies (e.g., Latour 1987). Kuhn's (1962) introduction of paradigm-led developments focused on the macro level (context of justification), and in recent years attention has shifted to the 'context of application', that is, the growing importance of the socio-economic environments of knowledge production (e.g., Nowotny et al. 2001). However, it is important to take into account all three contexts involved in the dynamics of knowledge because the resulting dynamics of search regimes involve these analytically distinct processes (Heimeriks 2009). For example, a field may be characterised by a strong and stable disciplinary identity in terms of publication patterns, while a diverging variety of skills and tools is used in research practices (e.g., genetics).

Furthermore, nations differ in terms of research portfolios. The dynamics of different search regimes are rarely taken into account when governance instruments are designed or evaluated. Most instruments for the governance of science and innovation are applied across several if not all fields in the natural sciences, social sciences, and humanities, and are intended to have the same effects in those fields.

Addressing these shortcomings, the current paper aims at developing a model of science as a complex adaptive system and using scientometric data to empirically investigate the emergence of search regimes from interactions among (i) the micro behaviour of researchers and (ii) emergent scientific fields within (iii) socio-economic environments. We assume an information-science perspective on knowledge dynamics and build on theoretical and methodological advances in the interdisciplinary fields of systems theory, scientometrics, and evolutionary economics.

Scientometric indicators are widely used in policy documents as well as in science, technology and innovation studies. To comprehend what the main features of science and innovation are, how they interrelate, and how these features and their relationships change is the chief purpose of scientometric indicators (e.g., Leydesdorff 2006).

The objective of this paper is to empirically disaggregate knowledge dynamics both in horizontal (field related) and vertical (context related) dimensions by articulating the three different dynamics and their path dependencies (in research, science, and society) in co-evolutions among one another leading to distinct search regimes in each field. This distinction of three dynamics will enable us to specify different micro-operations of the knowledge system because researchers (i) are geographically positioned and therefore locally embedded, (ii) can maintain socio-economic exchange relations, and (iii) learn from the resulting dynamics with reference to their positions and relations (Heimeriks 2009). The developments of the three analytically distinct dynamics are not a priori coordinated and may thus develop in some dimensions and at some places more than at others.

As cases for our empirical operationalisation of search regimes, we chose Biotechnology, Genomics, and Nanotechnology. These 'new lead sciences' are characterised by rapid growth, divergent dynamics, and new complementarities creating the need for wide-ranging cross-disciplinary competences. Bonaccorsi (2007) argues that European science is relatively weak (compared to the USA) in these fields characterized by high growth, high diversity, human capital, and institutional complementarities. Grasping the fruits of these emerging techno-sciences is an objective of many government priority programs in a knowledge-based and globalizing economy.

Life sciences such as Biotechnology and Genomics have received extensive investments from both the public and private sectors, because of their growing impact. New treatments and drugs, genetically modified foods, biologically controlled production processes, new materials, biologically based computing and many other applications are improving health, the environment, and industrial, agricultural, and energy production (OECD 2009). Biotechnology presents a more established field within the life sciences, with a history of strong university-industry collaborations (Leydesdorff and Heimeriks 2001), while Genomics emerged more recently in relation to strong government involvement around the genome projects (Propp and Moors 2009).

Nanotechnology is also likely to have major economic and social impacts in the years ahead. It may help further miniaturise information technology devices, resolve fundamental questions related to the immune system, accelerate advances in genomics, and contribute to the generation of renewable energy. Inventive activities in nanotechnology have risen substantially since the end of the 1990s although the share of nanotechnology in total patenting remains relatively limited (OECD 2009).

The paper is structured as follows. First, the model of science as a complex adaptive system is described. Next, the methods for obtaining data and indicators of knowledge dynamics in the fields of Biotechnology, Nanotechnology, and Genomics are presented. After presenting the results concerning each of the three contexts, the emergence of regimes is discussed. The last section contains the conclusions.

**Data and Methods**

In order to empirically investigate distinct search regimes in the selected fields, their knowledge dynamics will be disaggregated both in horizontal (field related) and vertical (level related) dimensions. First, the different dynamics of the fields and their path dependencies are discussed (in research, science, and society).Thereafter, we turn to the co-evolutionary nature of these dynamics and the emergence of regimes in the fields under study. The fields are delineated using aggregated journal-journal citation patterns (Leydesdorff and Cozzens 1993) for the period 1998-2008 (described below). Indicators are then developed to map the micro behaviour of the researchers within these fields, as well as the emergent field dynamics and the socio-economic interactions.

Within the first environment—the local research context—relevant processes include the interdependency that exists among researchers when producing knowledge through shared infrastructures, databases, international collaboration, and the number of contributors to scientific publications. Several analysts have used scientometric data to study patterns of research collaboration. De Solla Price (1986) noted increasing collaborative publishing as long ago as 1963. More recently, scientometric studies have shown an increase in international collaboration (e.g.,; Persson et al., 2004; Wagner, 2008).

Our data permit a detailed assessment of the rate of collaboration among institutions within the fields under study and its growth over time. We begin our examination of collaboration and the extent to which institutional resources are combined in research (Hicks and Katz 1996). These dynamics can be made visible by mapping the following proxies: a) the average number of authors per paper, b) the number of organisations that contribute to a field, c) the number of cities and countries where the authors are based (that is, the geographical distribution of knowledge production).

Each publication in the dataset contains one or more addresses that enable us to specify the geographical location of each university and industry and therefore derive information about local path dependent dynamics and collaboration patterns. This geographical information allows us to make a geographic mapping of the institutional addresses and their relations using Google Maps (Leydesdorff and Persson 2010).

When using scientific literature to model scholarly discourses, a research field can be operationalised as an evolving set of related documents. Each publication can be expected to contribute to the further development of the field at the research front. The specific use of title words can then be considered as a signature of the knowledge claim in the paper: new words and combinations of words can be expected to represent variation, while each paper is at the same time selectively positioned into the intellectual organization of a field using context-relevant references (Leydesdorff 1989; Lucio-Arias & Leydesdorff, 2009). The dynamics of the second environment, the emergent science system, thus can be made visible by mapping the following indicators (e.g., Van den Besselaar and Heimeriks 2006): a) stability of topics through time, b) the occurrence of new topics and the c) number of publications.

We use a method for delineating specialties as described elsewhere (Leydesdorff and Cozzens 1993). This method is based on a factor analysis of the journal-journal citations matrix of the core journals of a specialty. The point of departure is the selection of a seed journal representative for and central to the new specialty. For every year, we determine the relational citation environment of that journal, using a threshold of 1%. For the resulting set of journals we can make the journal-journal citation matrix, with the citing patterns as the variables. A factor analysis of this matrix results in factors consisting of journals that entertain similar citation patterns. The factor on which the seed journal has its highest loading can be considered as a representation of the field under study. The other factors provide information about the set of research fields that are related to the field under study.

The third level under study in this paper is provided by the societal environment in which the science system evolves. Science can be considered as an open communication system that is coupled to other parts of society; it is neither internally, nor externally determined, its development is caused by a complex interplay of internal and external factors; it is a relatively autonomous system. These societal dynamics can be made visible by mapping the non-academic references (patents, reports, etc.) which provide information about this context of application.

In this study, we focus on the so-called Triple Helix interactions: the participation of universities, companies, and governments in knowledge production. The Triple Helix model assumes the traditional forms of institutional differentiation among universities, industries, and government as its starting point. The model thus takes account of the expanding role of knowledge in relation to the political and economic infrastructure of the larger society (Etzkowitz and Leydesdorff 2000).

The following table (Table 1) summarizes the discussion of indicators at the three levels. For practical reasons, we limit ourselves in this study to publication-based indicators.

**Table 1**. Overview of the three levels of analysis and the indicators associated.

| Level of Analysis | Relevant Concepts | Relevant Processes | Indicator |
|---|---|---|---|
| Research | interdependency, complementarities, | intellectual division of labour, shared infrastructures, databases | geographical distribution, international co-authors, number of contributors |
| Science | size, disciplinarity | growth, convergence, divergence | number of publications, number of journals, emergence of new topics |
| Society | knowledge use in society, audience plurality and diversity | patenting, funding, research for policy, norms, etc | out of academia (co-) authors, patents, programmes, funding, non-academic output |

Interactions among the local research practices, scientific fields and society are multidirectional and involve positive and negative feedback loops (Figure 1).

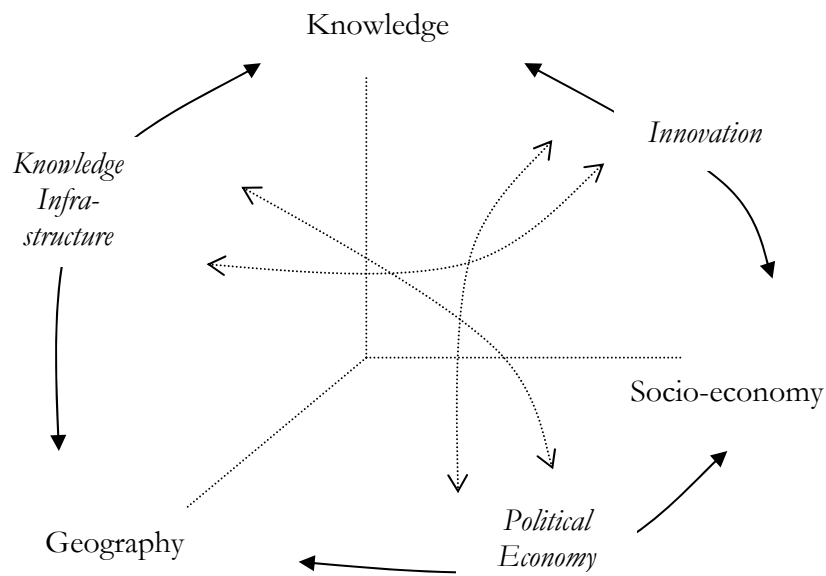

**Figure 1**. Three dynamics of search regimes.

In order to study the emergence of regimes, we first explore the relationship between the geography of knowledge production and topics using factor analysis. Factor analysis of all countries over the 500 most used title words provides a first indication of the geographical specificity of the topics under study. Additionally, we use entropy statistics for analysing emergent patterns. Probabilistic entropy is used here to study variety patterns in scientific developments. Though the concept of entropy originated from thermodynamic systems, it has acquired a general probabilistic meaning that allows for a large number of applications (Theil 1972; Grupp 1990). Entropy statistics is based solely on the properties of probability distributions, and, as such, is especially suitable for studying evolutionary phenomena at the level of any population of heterogeneous entities (Saviotti 1996).

In this study we use entropy statistics to study the distribution of topics (as indicated by the 500 most occurring title words) over all countries. The larger the entropy value, the larger the variety within a distribution of topics. The mutual information value equals zero when there exists no coupling/dependence between any of the two dimensions, and the higher the mutual information value the higher the degree of coupling. We are interested in the emergence of stable patterns (reproduced in time) in the three dimensions of i) geography, ii) knowledge, and iii) social and economical dynamics; we draw conclusions about the search regimes and the co-evolutionary dynamics giving rise to search regimes in the three fields under study.

**Results**

After obtaining the set of journals (described below in the section of science dynamics), all the publications for the period 1998-2008 were downloaded from the Web of Science. The total number of articles were 21,833 (genomics), 8,617 (biotechnology) and 84,044 (nanotechnology). In this period, nanotechnology showed the largest increase in the number of publications, from 4,696 in 1998 to 10,823 in 2008 (Table 2).

**Table 2**. Growth of number of publications in Genomics, Biotechnology and Nanotechnology.

| Field | Indicator | total | 1998 | 2008 | change % |
|---|---|---|---|---|---|
| genomics | $N$ | 21,833 | 1,619 | 2,526 | 56.02 |
|  | number of authors per paper | 5.27 | 4.69 | 6.03 | 28.63 |
|  | International co-authored % | 66.95 | 60.78 | 74.15 | 22.00 |
|  | number of countries | 97 | 51 | 64 | 25.49 |
| biotechnology | $N$ | 8,617 | 597 | 954 | 59.80 |
|  | number of authors per paper | 4.21 | 3.60 | 4.61 | 28.07 |
|  | International co-authored % | 57.31 | 43.89 | 62.26 | 41.88 |
|  | number of countries | 89 | 45 | 45 | 0.00 |
| nanotechnology | $N$ | 84,044 | 4,695 | 10,823 | 130.52 |
|  | number of authors per paper | 4.80 | 4.38 | 5.13 | 17.31 |
|  | International co-authored % | 64.51 | 57.06 | 67.98 | 19.13 |
|  | number of countries | 112 | 75 | 86 | 14.67 |

The first set of analyses focused on the geographically localised research practices. In all three fields, the average number of authors per paper showed a steady increase in the period under study. In genomics, the numbers increased from 4.69 to 6.03 authors per article on average, while in biotechnology this number increased from 3.60 to 4.61. Nanotechnology showed a relatively modest increase from 4.38 to 5.13.

All three fields are characterised by pronounced patterns of international collaboration as indicated by the number of articles with authors from more than one country. Genomics not only showed the highest proportion of internationally co-authored papers, but it also showed a high increase in international collaborations in the period under study: from 60.78% of the publications in 1998 to 74.15% in 2008.

Geographical distribution patterns show that the number of countries that contribute to biotechnology remains stable at 45 in the period under study, while nanotechnology and genomics show an increase in the number of countries participating in the development of the field. Although the number of countries and cities remained relatively stable in Biotechnology, the contributions from different parts of the world show a very dynamic development in the decade under study (Table 3).

**Table 3**. Share (%) of most important countries contributing to the field of Biotechnology in the period 1998-2008.

| Country | 1998 | 1999 | 2000 | 2001 | 2002 | 2003 | 2004 | 2005 | 2006 | 2007 | 2008 | Total |
|---|---|---|---|---|---|---|---|---|---|---|---|---|
| USA | 38.79 | 33.30 | 29.72 | 25.37 | 29.70 | 26.09 | 25.55 | 28.70 | 26.34 | 24.06 | 30.82 | 28.39 |
| GERMANY | 6.35 | 6.64 | 7.26 | 7.41 | 7.45 | 7.07 | 7.09 | 5.83 | 6.64 | 8.11 | 7.16 | 7.05 |
| JAPAN | 5.32 | 5.60 | 5.29 | 6.19 | 4.08 | 5.02 | 7.02 | 6.82 | 8.58 | 6.40 | 9.99 | 6.65 |
| SOUTH KOREA | 4.09 | 3.80 | 4.36 | 5.58 | 4.47 | 6.47 | 5.35 | 6.51 | 5.11 | 5.80 | 3.78 | 5.11 |
| ENGLAND | 4.81 | 5.79 | 5.64 | 5.41 | 5.33 | 5.28 | 4.28 | 4.46 | 4.17 | 4.31 | 4.32 | 4.78 |
| CANADA | 2.46 | 4.93 | 5.04 | 2.79 | 5.09 | 2.91 | 3.95 | 4.53 | 4.06 | 3.89 | 3.23 | 3.88 |
| NETHERLANDS | 4.61 | 3.42 | 5.21 | 6.10 | 5.72 | 4.29 | 3.61 | 2.48 | 2.70 | 2.23 | 3.18 | 3.74 |
| CHINA | 1.33 | 0.85 | 1.02 | 3.14 | 2.04 | 2.31 | 4.08 | 4.40 | 6.29 | 5.89 | 5.22 | 3.74 |
| SPAIN | 4.71 | 3.51 | 3.25 | 3.66 | 3.37 | 2.58 | 4.41 | 3.78 | 2.82 | 2.27 | 2.19 | 3.18 |
| ITALY | 2.35 | 1.42 | 2.73 | 2.27 | 2.59 | 4.23 | 3.21 | 3.22 | 2.82 | 2.97 | 2.34 | 2.80 |
| FRANCE | 3.58 | 5.98 | 3.25 | 2.70 | 2.98 | 3.96 | 3.21 | 2.17 | 1.76 | 1.62 | 1.84 | 2.79 |
| SWITZERLAND | 3.28 | 2.37 | 2.31 | 2.79 | 3.29 | 3.04 | 2.07 | 2.36 | 2.23 | 2.78 | 1.89 | 2.54 |
| TAIWAN | 1.54 | 2.09 | 1.88 | 2.01 | 1.57 | 3.17 | 2.61 | 2.67 | 2.29 | 3.66 | 2.49 | 2.48 |
| SWEDEN | 1.54 | 2.66 | 3.67 | 2.62 | 2.90 | 1.78 | 3.01 | 2.05 | 2.59 | 1.48 | 2.24 | 2.35 |
| INDIA | 1.23 | 1.90 | 1.28 | 2.27 | 2.66 | 2.71 | 2.14 | 1.12 | 2.18 | 1.76 | 1.19 | 1.84 |

In summary, knowledge production in Biotechnology seems to shift away from the USA towards Asian regions (most importantly Seoul, Tokyo, Beijing, and Singapore). European knowledge production in Biotechnology remains rather stable compared to the USA. Nanotechnology and Genomics show equally pronounced developments. Genomics was initially dominated by a small group of countries (most importantly, the USA). In later years many countries from all over the world increased their scientific output in the field. These developments reflect the process of ongoing globalisation and the consequent escalation in scientific competition (Wagner 2008).

**Science dynamics**

The previous section discussed the local research dynamics in the fields under study. After having taken these developments into account, a further question deals with the developments at the level of the emergent scientific landscapes. At the science level the three fields show very distinct patterns of development.

Nanotechnology and Genomics exhibit an increase in the number of journals that together form the field. In Nanotechnology, the journals Advanced Materials, Nanotechnology, Applied Physics Letter, Journal of Applied Physics and Nano Letter form a cluster in 2008. The field Genomics in 2008 is composed of Genomics Research, P Natl Acad Sci USA, Nucleic Acid Research, BMC Genomics, Bioinformatics and Genome Biology.

Biotechnology, the oldest of the fields under study, shows a different pattern of development. In recent years, it seems that there is a narrow definition of the field consisting of only the three journals: Biotechnology Progress, Biotechnology Bioengineering, the Journal of Biotechnology, and a broader definition interwoven with applications of microbiology.

The use of title words in the fields under study provides us with an indication of the cognitive developments within the field. The graph below provides a selection of the most important title words in Genomics.

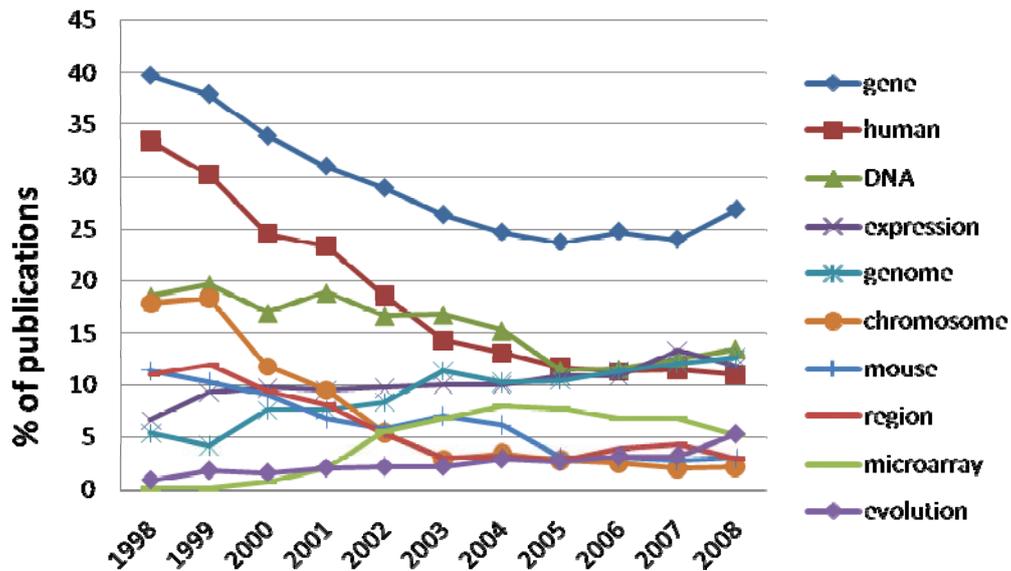

**Figure 2**. The development of title words in genomics.

Genomics shows much more shared use of title-words than Biotechnology and Nanotechnology, indicating more common topics within that field. Especially in 1998 and 1999, a large proportion of the research was focused on the human genome as more than 30 percent of the papers used the title words 'gene' and 'human'. However, the level of stability in title words dropped rapidly in a period of turbulent growth in the number of publications. In contrast, Biotechnology shows increasing usage of shared title words such as 'cell' and 'protein'. Figure 3 shows some of the most frequent title words, as well as some examples of topics that almost disappeared in the period under study.

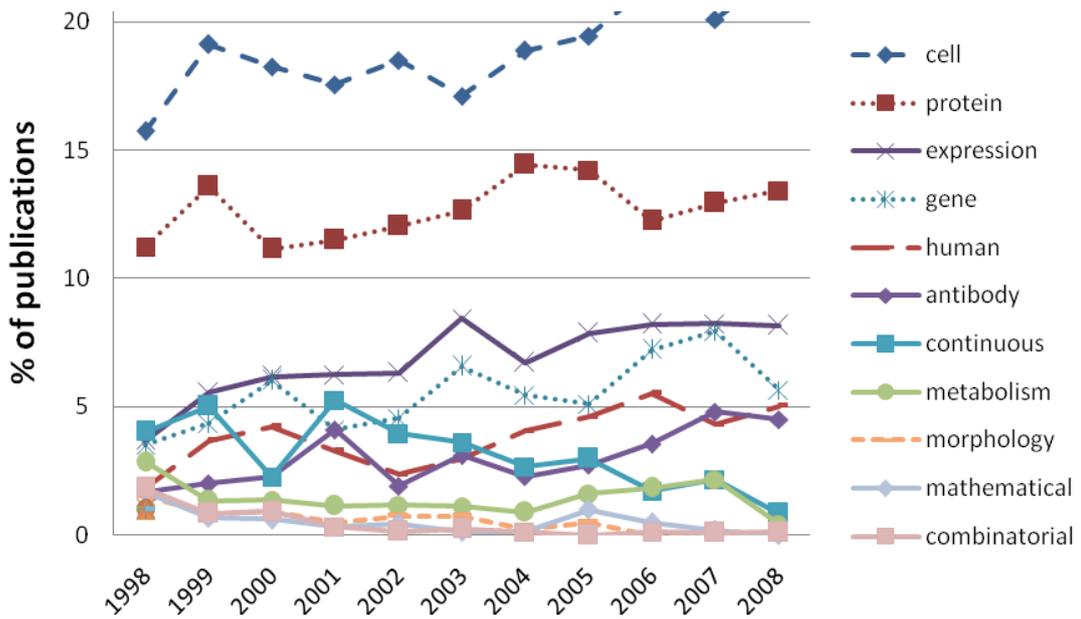

**Figure 3**. The development of title words in Biotechnology.

Nanotechnology, the largest of the three fields under study and with the fastest growth-rates still shows a higher level of shared title words than Biotechnology in the years 1998 and 1999.

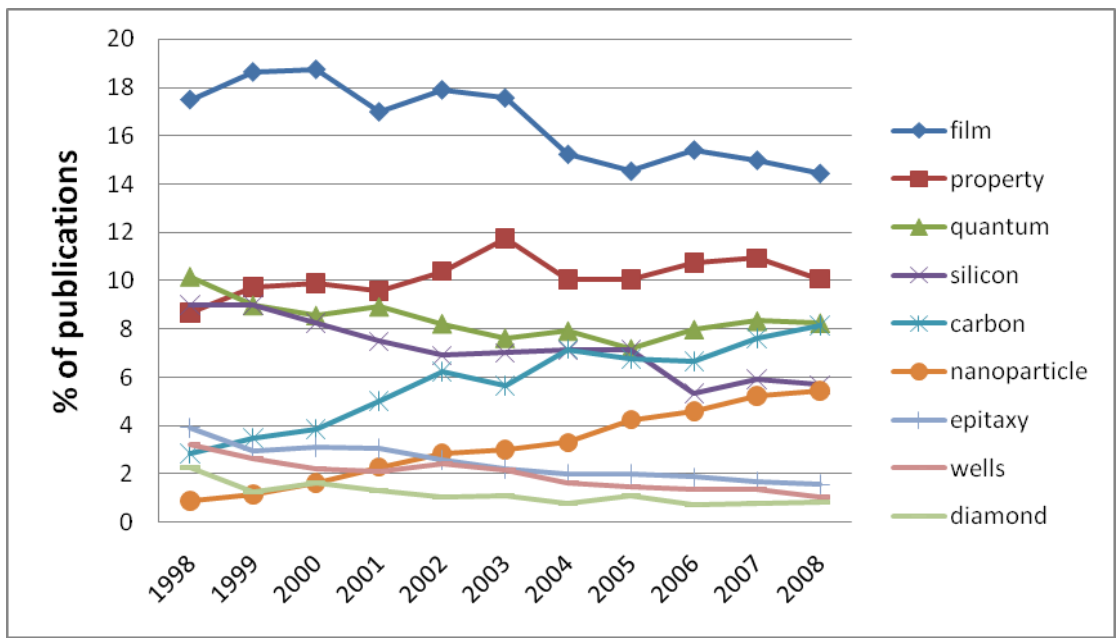

**Figure 4**. The development of title words in Nanotechnology.

In summary, the potential set of topics that is available to researchers is different among the three fields under study. Especially in Nanotechnology the divergent variety of topics provides a wide set of possibilities to add new knowledge claims to the body of knowledge.

**Societal dynamics**

In a knowledge-based society, most societal problems require new knowledge developments (Webster 2006), and thus 'problem-based' R&D will become increasingly important. In this study, these societal dynamics are made visible by the contribution of researchers outside universities to knowledge production. Involvement of non-academics may reduce the strength of scientific entry barriers in a field, while the contribution of exclusively academic researchers may decrease the appropriability of knowledge.

**Table 4**. Relative contributions of universities, industry and public-sector laboratories in the period 1998-2008.

| Field | Type of organisation | Average 1998-2008 | 1998 | 2008 |
|---|---|---|---|---|
| genomics | Public sector | 28.24 | 29.84 | 26.00 |
|  | Industry | 3.67 | 3.20 | 3.29 |
|  | University | 61.09 | 58.53 | 63.59 |
|  | not-classified | 7.00 | 8.43 | 7.12 |
| biotechnology | Public sector | 20.12 | 21.45 | 17.94 |
|  | Industry | 5.54 | 4.60 | 5.47 |
|  | University | 64.74 | 65.99 | 67.74 |
|  | not-classified | 9.59 | 7.97 | 8.85 |
| nanotechnology | Public sector | 25.80 | 27.36 | 24.50 |
|  | Industry | 5.04 | 6.75 | 3.82 |
|  | University | 62.49 | 58.69 | 65.15 |
|  | not-classified | 6.67 | 7.20 | 6.53 |

An important indication of the societal context in which the fields develop is provided by the so-called Triple Helix interactions: the participation of universities, companies and government agencies in knowledge production. In all three fields, the relative importance of public-sector laboratories has declined in this period of rapid growth between 1998-2008. Biotechnology shows an increase in industrial participation in knowledge production, Genomics shows a stable pattern and Nanotechnology a significant decline in the number of companies publishing in the field. This development points at a widening gap between (academic) research and a commercial context of application in Nanotechnology. Possibly, corporations that were previously quite active in research started to reduce or outsource their research to universities or to specialised organisations (OECD 2005).

**Emerging regimes**

After presenting the results concerning each of the three contexts, we now turn to the emergence of regimes. As mentioned, interactions between the local research practices, scientific fields, and socio-economical contexts are multidirectional and can be expected to involve positive and negative feedback loops. In other words, research, science, and society interact and shape one another in processes of co-evolution (Whitley 2000; Rip 2002).

In the previous section we showed (Table 3), how more countries and cities contribute to knowledge production in the new leading sciences in a process of ongoing globalisation and scientific competition (Wagner 2008). In parallel to the increased globalisation, we have shown the different patterns of turbulent cognitive development in the fields under study. A scientific field ('body of knowledge') constrains the set of trajectories that a researcher can explore, as well as the range of available strategies, competencies, and forms of organisation. Science relates to research activities insofar as the science level creates resources for individual researchers, such as recognition and reputation, which feed back into their research practices over time (Rip 1990: 389).

At the research level, the final outcome of research efforts has to be contextualised, written, and edited. In these local actions, researchers respond to the emergent science level (in the form of existing and expected bodies of knowledge) in an anticipatory mode in which the existing claims in the body of knowledge are partially deconstructed and reconstructed, but also accepted to a large extent (Fujigaki 1998; Leydesdorff, 2010).

The relationship between the geography and topics were further explored using factor analysis. Biotechnology shows a more geographically distributed pattern than Nanotechnology and Genomics. The largest component resulting from factor analysis of all countries over the 500 most used title words explains 28.26% of the variance compared to 35.26% in Genomics and 32.75% in Nanotechnology.

The path-dependent nature of knowledge production implies that different countries and research organisations are likely to find different local optima (Frenken et al. 2009). In some cases the emergent knowledge base is such that researchers are compelled to explore the same set of cognitive, technological, and methodological resources and to adopt the same search procedures. In other cases, the emergent knowledge base instead allows researchers to pursue different behaviours. The possibility to exploit emerging opportunities is determined by the need to co-ordinate due to interdependencies among researchers (Whitley 2000).

Using entropy statistics is a way to measure to degree of diversity and the mutual information among the three contexts of research, science, and society. Concerning the frequency distributions of topics and geography, the (Shannon) H values inform us that an overall increase in variety occurred in the period under study (Figure 5). The transmission values (T) inform us about the mutual information content between the dimension 'country' on one hand, and the dimension 'research topic' on the other hand, as represented by the title-words. All countries in each set were used in combination with the 500 most occurring title words. Nanotechnology shows the largest variety within a distribution of topics over countries (Figure 5).

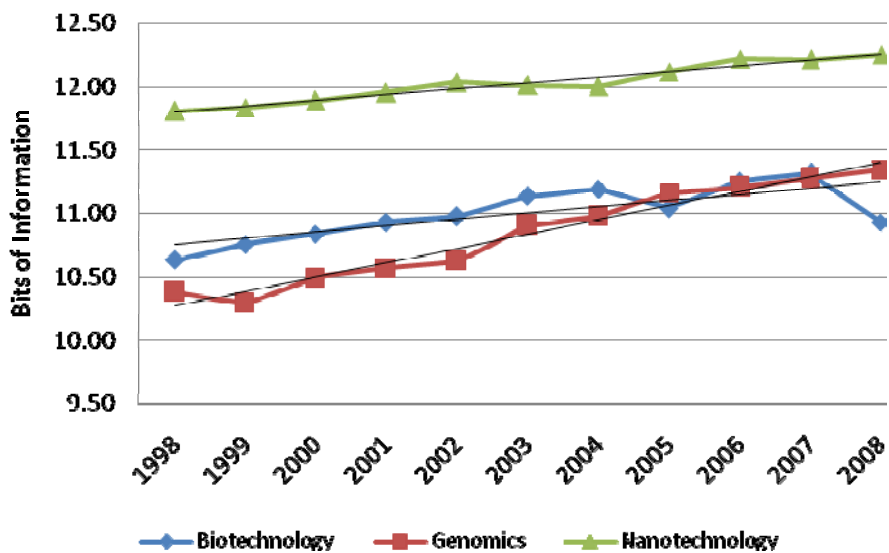

**Figure 5**. The two-dimensional entropy values (*H*) of all countries and most important topics in the period under study.

The transmission values show very different patterns in the three fields under study (Figure 6). In Biotechnology the transmission values are relatively high and indicate a strong relationship between the geography of knowledge production and the research topics under study. This result is in line with the results of the factor analysis discussed above that suggested a connection between geography and topics under study.

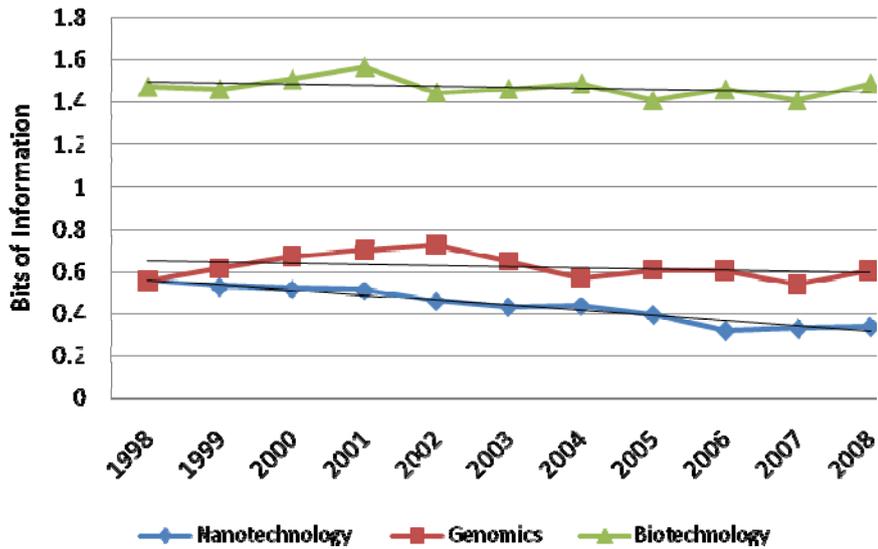

**Figure 6**. The transmission values between all countries and most important topics.

In Nanotechnology and Genomics, the transmission values are much lower than in Biotechnology, indicating a weaker relationship between the topics of research and the geography of knowledge production. In Nanotechnology, this process of de-coupling is still ongoing more than in Genomics. Biotechnology is much more rooted in local contexts, possible related to socio-economical contexts of application. Nanotechnology shows the highest entropy values and the lowest transmission values. This suggests that while there is a divergent set of options available for researchers around the globe to contribute in this field, the local contexts are less relevant in knowledge production.

The distributions of organisation type (academic, public sector, and commercial) over countries shows a pattern of convergence; in all three fields the transmission values between countries and organisation types decreases. In Nanotechnology the decrease is most pronounced, indicating that initially public sector and commercial organisations were unevenly represented in the field across countries (Figure 7).

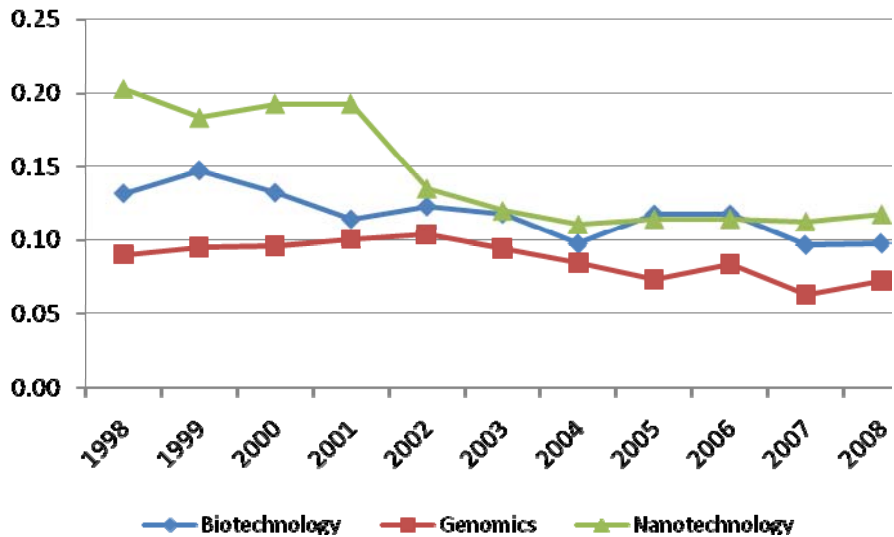

**Figure 7**. The transmission values between all countries and type of organisation.

The transmission values between word use and organisation type show that in Nanotechnology a sharp decrease occurs between 1998 and 2008. This development coincides with the surge in funding of Nanotechnology when it became a priority funding area in most advanced nations in the period 2000-2003 (Leydesdorff and Schank 2008). Previously, different organisation types contributed to distinct topics of research in this field while in later years this specificity became much less pronounced. Biotechnology maintains a fairly stable pattern of institutional specificity with respect to the topics of research (Figure 8).

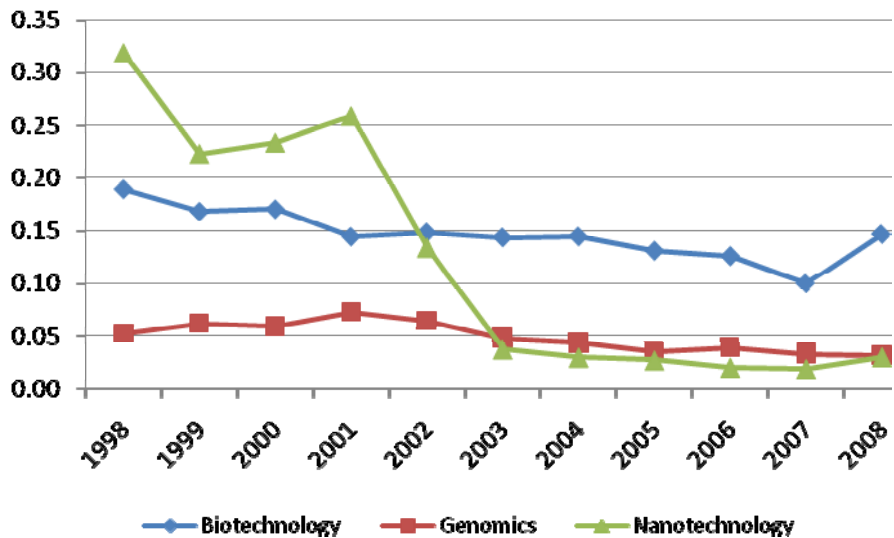

**Figure 8**. The transmission values between type of organisation and most important topics.

In summary, research fields exhibit institutional and localised knowledge dynamics that give rise to distinct regimes in a co-evolution with the scientific bodies of knowledge and the contexts of application. A search regime can be considered as an evolving configuration among the dynamics of the different contexts. The evolutionary regimes are expected to remain in transition because the differentiation in the codification (among research practices, scientific communication, and socio-economical valorisation) generates a feedback that changes the organization and dynamics of the search regimes (Leydesdorff 2006). The three dynamics are treated as equivalent in the model, but they are substantially very different. The selection mechanisms are expected to operate asymmetrically. Regimes can thus be distinguished in terms of the extent to which a synergy is self-organized among the main subdynamics.

Our study showed that the mutual dependency of researchers is high in the new leading sciences in the period under study. Mutual dependency is reflected in the collaboration patterns of researchers. Especially in Genomics, the need to collaborate is obvious; the number of authors numbers increased to 6.03 authors per article on average, and large majority of the contributions were internationally co-authored.

Researchers think globally and act locally; the scientific field constrains the set of trajectories that a researcher may explore. Science thus relates to research activities insofar as the science level creates opportunities for researchers which feed back into their research practices over time. The scope of opportunities for researchers around the world to contribute within the constraints of the existing body of knowledge were different in each field. Especially in Nanotechnology the divergent variety of topics provides newcomers possibilities to add new knowledge claims.

Additionally, the context of application provides path dependent dynamics for knowledge production and dissemination since cognitive, social and geographical proximity are crucial for socio-economic valorisation (e.g. Boschma and Frenken 2009). Biotechnology showed the highest level of specialisation with respect to different local contexts in this study. The specialised role of public sector and commercial organisations was initially highest in Nanotechnology, but this division of labour disappeared as the field matured.

**Policy implications**

As a consequence of increasing globalisation, an competition, there has been a growing emphasis on science and innovation in industrialised countries (e.g. Cowan et al. 2000). Governments, both nationally and regionally, need to ensure that the local knowledge base is strong and therefore attractive to globally competitive companies (Foray 2006).

The analyses presented here have major implications for research and innovation policy. The innovation systems literature emphasises that because science and innovation are locally embedded in complex systems, practises in research and innovation policies cannot be simply copied between countries and fields (Asheim et al. 2006; Bonaccorsi 2007). Research fields exhibit distinct and localised knowledge dynamics that respond differently to government interventions. Conceptualisations of inter-science differences and dynamics are thus increasingly important.

Bonaccorsi's (2007) model of search regimes provides a useful starting point. Nevertheless, there are a number of questions with the Bonaccorsi framework that need to be addressed. For example, what constitutes a regime? How does a scientific field switch from one category to another? What might bring about such a change?

Just focusing on horizontal disaggregation (differences among fields) does not allow for accurate policy intervention. It is important to take into account the three different levels of analysis because the dynamics used to characterise search regimes relate to different processes in different selection environments. Existing models of science insufficiently address these different levels of analysis of knowledge dynamics. Following Simon (1973), we argue that in addition to horizontal disaggregation, we need to take into account vertical disaggregation of knowledge dynamics to understand the dynamics of search regimes.

The regimes are systemic, that is, largely beyond (government) control; further developments are based on the self-organization of the interactions among the contexts of discovery, justification and application. The subdynamics can also be considered as different sources of variance which disturb and select from one another (Leydesdorff 2006).New methods, tools and collaboration patterns are continuously introduced in research practices (Heimeriks and Vasileiadou 2008), the landscape of scientific publications is continuously in flux (Leydesdorff and Cozzens 1993) and new applications are being developed at any given moment (Nelson 1994).

Resonances among selections can shape trajectories in co-evolutions, and the latter may recursively drive the system into new regimes (Dolfsma and Leydesdorff 2009). Our analyses show that the role of non-academic organisations in knowledge production differs among fields, as well as the variety of topics that are (potentially) available to researchers. Furthermore, the need to collaborate and the entry barriers for newcomers are different among fields of knowledge production. To increase the probability of policy success, research and innovation policies need to account for the different contexts that provide opportunities but also set limits to what can be achieved by policy. Doing so, public intervention should neither apply 'one-size-fits-all' frameworks nor adopt 'picking-winner' policies. Well informed governance is needed to understand the opportunities in the knowledge-based economy and to construct unique locational advantages in relation to the global body of knowledge and the societal dynamics.

**Conclusions**

It is expected that the fields Biotechnology, Genomics and Nanotechnology will be important drivers of productivity and economic growth, as well as a vital resource in addressing societal challenges. Consequently, these fields are national priority areas in almost all countries. Each of these 'new leading sciences' is currently growing at a rapid rate and experiencing divergent dynamics of search. According to Bonaccorsi (2008), this situation is in sharp contrast to more traditional scientific fields that are characterised by convergent patterns of slow growth.

Furthermore, new forms of complementarities arise, in the form of processes of collaborative competence building and institutional cooperation across different types of actors. These developments were made visible as the increasing (international) collaborations as well as the consistent public and commercial involvement in the fields.

Although the selected fields are all be characterised by rapid growth and divergent dynamics of search, we have shown some important differences among these emergent search regimes. These differences are most visible in the emergence of search regimes as resulting from the interaction of the three sources of variance. The regimes can be distinguished in terms of the extent to which a synergy is self-organized among the three (analytically distinguishable) subdynamics.

Our study showed that the scope of opportunities for researchers around the world to contribute within the constraints of the existing body of knowledge were different in each field. Genomics was characterised by relatively low variety in topics across the globe compared to Biotechnology and Nanotechnology. Especially in Nanotechnology the divergent variety of topics can be expected to provide newcomers possibilities to add new knowledge claims.

Additionally, the relevance of a local context of application contributes to the knowledge dynamics to various degrees in the fields under study. Biotechnology showed the highest level of geographical specialisation with respect to different local contexts in this study. The specialised role of public sector and commercial organisations was initially highest in Nanotechnology, but this division of labour disappeared as the field matured. These results suggest that the surge in funding in the period under study contributed to an intellectual reorganisation of the field of Nanotechnology.

Interest is growing in improving the understanding of how the sciences evolve and stimulate the growth and competitiveness of economies, and contribute to societal challenges. This paper contributes to improving research and innovation indicators by combining existing path dependencies at the levels of research, science, and society. We have shown that regimes emerge from interactions among local research practices, emergent scientific landscapes, and the field's relationship to its societal context. Furthermore, the nature of these regimes is different in each field of knowledge production.

As the number of countries contributing to knowledge production increases, these insights become increasingly relevant for public funding of science and innovation.

Public funding has been implemented predominantly in the context of nations. The growth of the global network of science means that nations must take careful stock of the conduct of science at the global level as well as at the national and regional levels, in relation to locally existing absorptive capacity. Research governance thus entails a linking and sinking strategy as proposed by Wagner (2008). It links to global science dynamics and locally 'sinks; efforts by taking into account local research dynamics with respect to stakeholders, infrastructures and the local knowledge base in terms of human resources and skills.

Furthermore, the regime approach to knowledge dynamics means that one can appreciate a variety of relevant spaces for public intervention, since some regimes require international research and innovation policies while others are the realm of regional policies. This means that the location of new research programmes and the geography of scientific knowledge production more broadly, are subject to path-dependent dynamics where research programmes may prosper in some locations and become marginalized in other locations.